\documentclass[epj,nopacs]{svjour}
\usepackage{epsfig,amsmath}
\usepackage{graphicx}

\newcommand{\nn}{\nonumber}
\newcommand{\eps}{\epsilon}
\newcommand{\Mpl}{\overline{M}_{\rm Pl}}
\newcommand{\qs}{\!\!\not\!q}

\begin{document}

%%%%%%%%%%%% Begin Cover Page %%%%%%%%%%%%%%%%%%%%%%%%%%%%%%%%%%%%%%%%%%
\title{HELAS and MadGraph with spin-3/2 particles}

\author{
 K.~Hagiwara\inst{1}
 \and 
 K.~Mawatari\inst{2,3}\fnmsep\thanks{e-mail: 
            kentarou.mawatari@vub.ac.be}
 \and 
 Y.~Takaesu\inst{1}\fnmsep\thanks{e-mail: takaesu@post.kek.jp} }
\institute{
 KEK Theory Center, and Sokendai, Tsukuba 305-0801, Japan 
 \and 
 Theoretische Natuurkunde and IIHE/ELEM, Vrije Universiteit Brussel,\\
 and International Solvay Institutes,
 Pleinlaan 2, B-1050 Brussels, Belgium
  \and
 Institut f\"ur Theoretische Physik, Universit\"at
 Heidelberg, Philosophenweg 16, D-69120 Heidelberg, Germany}

\abstract{Fortran subroutines to calculate helicity amplitudes with
massive spin-3/2 particles, such as massive gravitinos, 
which couple to the standard model and supersymmetric
particles via the supercurrent, are added to the {\tt HELAS} 
({\tt HEL}icity {\tt A}mplitude {\tt S}ubroutines) library. They are
coded in such a way that arbitrary amplitudes with external gravitinos 
can be generated automatically by {\tt MadGraph}, after slight
modifications. All the codes have been tested carefully by making use of 
the gauge invariance of the helicity amplitudes.}

\titlerunning{HELAS and MadGraph with spin-3/2 particles}
\authorrunning{K.~Hagiwara, K.~Mawatari and Y.~Takaesu}

\maketitle
%%%%%%%%%%%% End of Cover Page %%%%%%%%%%%%%%%%%%%%%%%%%%%%%%%%%%%%%%%%%

%%%%%%%%%%%%%% Begin Main Part %%%%%%%%%%%%%%%%%%%%%%%%%%%%%%%%%%%%%%%%%

\vspace*{-105mm}
\noindent KEK-TH-1417\\
\noindent HD-THEP-10-20%\\
%\today
\vspace*{80mm}

%%%%%%%%%%%%%% Begin Section 1 %%%%%%%%%%%%%%%%%%%%%%%%%%%%%%%%%%%%%%%%% 
\section{Introduction}\label{intro}

Gravitinos are spin-3/2 superpartners of gravitons in local 
supersymmetric extensions to the Standard Model (SM). 
If supersymmetry (SUSY) breaks spontaneously, 
gravitinos absorb massless spin-1/2 goldstinos and become massive 
by the super-Higgs mechanism. 
Therefore, the gravitino mass is related to the scale of SUSY
breaking as well as the Planck scale like
\begin{align}
 m_{3/2}\sim (M_{\rm SUSY})^2/M_{\rm Pl}.
\end{align}
This implies that the gravitino can take a wide range of mass,
depending on the SUSY breaking scale, from eV up to scales beyond
TeV, and provide rich phenomenology in particle physics
as well as in cosmology~\cite{Giudice:1998bp}.  

Although the gravitino can play an important role even in collider
signatures when it is the lightest supersymmetric particle (LSP),
there is few Monte Carlo event generators which can treat them.%
\footnote{The SM with gravitino and photino is supported by 
{\tt WHIZARD}~\cite{Kilian:2007gr}.} 
In this paper, we present new {\tt HELAS} 
subroutines~\cite{Hagiwara:1990dw} for the massive
gravitinos and their interactions based on the effective Lagrangian below, 
and implement them into
{\tt MadGraph/MadEvent (MG/ME) v4}~\cite{Stelzer:1994ta,Maltoni:2002qb,Alwall:2007st}
so that arbitrary amplitudes with external gravitinos 
can be generated automatically.%
\footnote{The Fortran code for simulations of the massive gravitinos
is available at the KEK HELAS/MadGraph/MadEvent Home Page,
{\tt http://madgraph.kek.jp/KEK/}.}

The effective interaction Lagrangian relevant to the gravitino
phenomenology is~\cite{Wess:1992cp,Moroi:1995fs,Bolz:2000fu} 
\begin{align}
  {\cal L}_{\rm int}
 =&-\frac{i}{\sqrt{2}\,\Mpl} \nn\\
   &\ \times\big[\bar{\psi}_{\mu}
    \gamma^{\nu}\gamma^{\mu}P_{L}f^i\,
    (D_{\nu}^{}\phi^i_{L})^*-\bar{f^i}P_{R}
   \gamma^{\mu}\gamma^{\nu}\psi_{\mu}\,
   (D_{\nu}^{}{\phi}^i_{L})  \nn\\
  &\ \ -\bar{\psi}_{\mu}
    \gamma^{\nu}\gamma^{\mu}P_{R}f^i\,
    (D_{\nu}^{}\phi^i_{R})^*+\bar{f^i}P_{L}
   \gamma^{\mu}\gamma^{\nu}\psi_{\mu}\,
   (D_{\nu}^{}{\phi}^i_{R})\big] \nn\\
  &-\frac{i}{8\Mpl} 
   \bar{\psi}_{\mu}[\gamma^{\nu},\gamma^{\rho}]
   \gamma^{\mu}\lambda^{(\alpha)a}F_{\nu\rho}^{(\alpha)a},
\label{L_int}
\end{align}
where $\psi^{\mu}$ is the spin-3/2 gravitino field, $f^i$ and $\phi^i$
are spinor and scalar fields in the same chiral supermultiplet,  
$P_{R/L}=\frac{1}{2}(1\pm\gamma_5)$ is the chiral-projection operator,
and
$\Mpl\equiv M_{\rm Pl}/\sqrt{8\pi} %=1/\sqrt{8\pi G_N}
 \sim 2.4\times 10^{18}$ GeV is the reduced Planck mass.
The covariant derivative is 
\begin{align}
 D_{\mu}^{}= \partial_{\mu}^{}+ig_{s}T^{a}_3A_{\mu}^{a}
            +igT^a_2W_{\mu}^a+ig'YB_{\mu},
\end{align}
where $g_s$, $g$ and $g'$ are the $SU(3)_C$, $SU(2)_L$ and $U(1)_Y$
gauge couplings, respectively, and
$T^a_3$, $T^a_2$ and $Y$ are the generators of the
$SU(3)_C$ $(a=1,\cdots,8)$, $SU(2)_L$ $(a=1,2,3)$ and $U(1)_Y$ groups.
The field-strength tensors for each gauge group are
\begin{align}
 F_{\mu\nu}^{(3)a}&= \partial_{\mu}^{}A_{\nu}^{a}
  -\partial_{\nu}^{}A_{\mu}^{a}-g_sf_3^{abc}A_{\mu}^{b}A_{\nu}^{c}, \\
 F_{\mu\nu}^{(2)a}&= \partial_{\mu}^{}W_{\nu}^{a}
  -\partial_{\nu}^{}W_{\mu}^{a}-gf_2^{abc}W_{\mu}^{b}W_{\nu}^{c}, \\
 F_{\mu\nu}^{(1)a}&= \partial_{\mu}^{}B_{\nu}^{}
  -\partial_{\nu}^{}B_{\mu}^{},
\end{align}
and the corresponding gauginos $\lambda^{(\alpha=3,2,1)a}$ are gluinos
($\tilde g^a$), winos ($\tilde W^a$) and bino ($\tilde B$), respectively.

The paper is organized as follows:
In Sect.~\ref{sec:sample} we give sample numerical results.
Sect.~\ref{sec:summary} presents our brief summary.
In App.~\ref{sec:helas_new} we give the new {\tt HELAS} subroutines
for spin-3/2 particles, and in App.~\ref{sec:mg} we   
describe how to implement the amplitudes into {\tt MG}.

%%%%%%%%%%%%%% Begin Section 2 %%%%%%%%%%%%%%%%%%%%%%%%%%%%%%%%%%%%%%%%% 
\section{Sample results}\label{sec:sample}

In this section, we present some sample numerical results, 
using the new {\tt HELAS} subroutines, 
which are presented in Appendix~\ref{sec:helas_new}, 
and the modified {\tt MG},
which is described in Appendix~\ref{sec:mg}.

In the gauge mediated SUSY breaking scenarios, the gravitino
is often the LSP, and its phenomenology depends on what is 
the next-to-lightest supersymmetric particle (NLSP).
Here we consider the stau NLSP scenario as well as the neutralino NLSP one.

\subsection{Stau NLSP}

\begin{figure}
 \centering 
 \epsfig{file=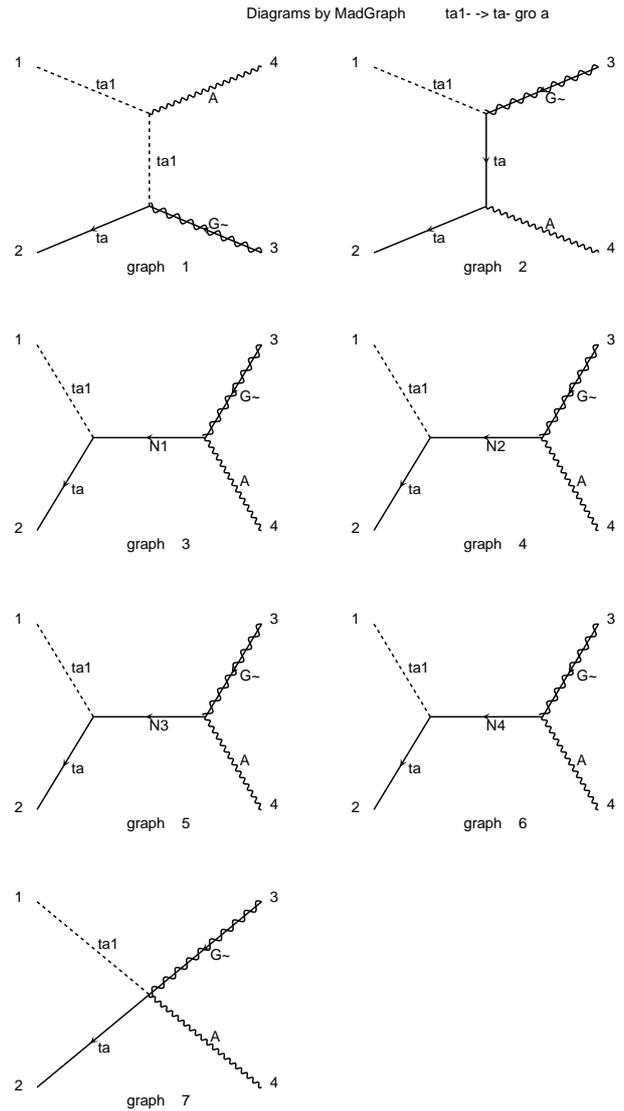,width=0.9\columnwidth,clip}
 \caption{Feynman diagrams for the radiative stau decay process
 for the $\tilde G$ LSP case,
 $\tilde{\tau}\rightarrow\tau\tilde{G}\gamma$, generated by
 {\tt MadGraph}. {\tt ta1}, {\tt ta}, {\tt G$\sim$}, {\tt A}, and {\tt Ni} 
 denote a stau, a tau-lepton,
 a gravitino, a photon, and neutralinos, respectively.}
 \label{diagrams}
\end{figure}

\begin{figure}
 \centering 
 \epsfig{file=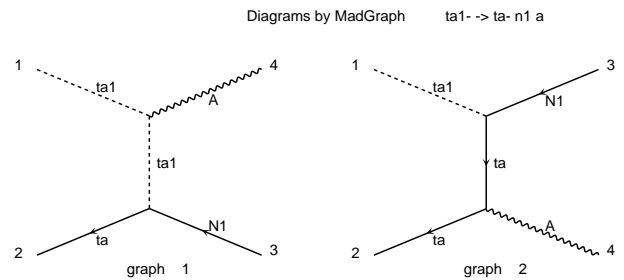,width=0.9\columnwidth,clip}
 \caption{The same as Fig.~\ref{diagrams}, but for the
 $\tilde{\chi}^0_1$ LSP case,
 $\tilde{\tau}\rightarrow\tau\tilde{\chi}^0_1\gamma$.}
 \label{diagrams_n}
\end{figure}

As a sample result for the stau NLSP scenario,
we consider radiative $\tilde\tau$ decays,
\begin{subequations}
\begin{align}
 \tilde{\tau}_{R}^-\rightarrow\tau^-\,\tilde{G}\,\gamma.
\label{staudecayprocess}
\end{align}
Here we regard the stau as a purely right-handed stau for simplicity.
Feynman diagrams shown in Fig.~\ref{diagrams} and the corresponding
helicity amplitudes
are generated automatically by the modified {\tt MG}.
To study the spin-3/2 nature of the gravitino, we compare the
$\tilde G$ LSP case 
\eqref{staudecayprocess} with the $\tilde\chi^0_1$ LSP case, 
\begin{align}
 \tilde{\tau}_{R}^-\rightarrow\tau^-\,\tilde\chi^0_1\,\gamma,
\label{staudecayprocess_n}
\end{align}
\end{subequations}
where only two decay diagrams contribute;
see Fig.~\ref{diagrams_n}.


We evaluate the amplitudes for the both cases, 
\eqref{staudecayprocess} and \eqref{staudecayprocess_n},
in the $\tilde\tau$ rest frame as
\begin{align} 
 p_{\tilde{\tau}} &= (m_{\tilde{\tau}},0,0,0),\nn\\
 p_{\gamma} &= (E_{\gamma},0,0,E_{\gamma}),\nn\\
 p_{\tau} &= (E_{\tau},p_{\tau}\sin\theta,0,p_{\tau}\cos\theta),\nn\\
 p_{\rm LSP} &= (E,p^{x},0,p^{z})\label{p4},
\end{align}
where the $z$-axis is taken along the photon momentum direction, 
and the $y$-axis is along 
$\overrightarrow{p}_\gamma\times\overrightarrow{p}_\tau$, the normal of 
the decay plane.

Using the generated helicity amplitudes and the above kinematical variables, 
we investigate photon polarizations by means of
Stokes parameters, $P_1, P_2$, and $P_3$, which are related with
the photon density matrix as
\begin{align}
 \frac{d\rho_{\lambda\lambda'}}{dE_\gamma\,d\cos\theta} =
 \frac{1}{2}
 \left(1+\sum^3_{i=1}P_i\sigma_i\right)_{\lambda\lambda'}\cdot
 \frac{d\Gamma_{\rm sum}}{dE_\gamma\,d\cos\theta}
\end{align}
with the Pauli sigma matrices $\sigma_i$.
$d\Gamma_{\rm sum}=d\rho_{++}+d\rho_{--}$ is the usual spin-summed 
differential decay rate.
The density matrix is calculated as 
\begin{align}
 d\rho_{\lambda\lambda'}=\frac{1}{2m_{\tilde\tau}}
 \sum{\cal M}^{}_\lambda{\cal M}^*_{\lambda'}\,d\Phi_3,
\end{align}
where ${\cal M}_\lambda$ is the helicity amplitude with the photon helicity
$\lambda$, and $d\Phi_3$ is the three-body phase space factor.
The summation symbol implies the summation over the tau and
gravitino/neutralino helicities. 
By definition, Stokes parameters take real values from $-1$ to $1$, and 
$P_3$ shows the right-left asymmetry of circular polarizations, 
while $P_1$ and $P_2$ present linear polarizations, which 
reflect the interference between the amplitudes for the 
right- and left-handed photons.

\begin{figure}
 \centering 
 \epsfig{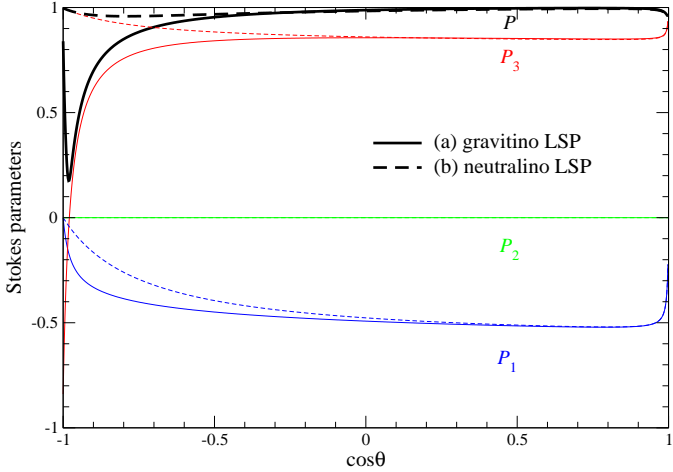}
\caption{Angular dependence of the Stokes parameters of the radiated photon
for the $\tilde\tau$ decay process,
$\tilde{\tau}_R\rightarrow\tau\tilde G\gamma$ (a) and
 $\tilde{\tau}_R\rightarrow\tau\tilde\chi_{1}^0\gamma$ (b),
where $\theta$ is the decay angle between the photon and the tau-lepton. 
We set $m_{\tilde\tau}=150$ GeV,  $m_{\rm LSP}=75$ GeV and $E_\gamma= 40$ GeV.} 
\label{fig:stau}
\end{figure}

In Fig.~\ref{fig:stau},
we show the $\cos\theta$ dependence of the Stokes parameters
of the radiated photon for $\tilde{\tau}_R\rightarrow\tau\tilde G\gamma$ (a)
and $\tilde{\tau}_R\rightarrow\tau\tilde{\chi}_1^0\gamma$ (b),
where we use 
\begin{align}
 m_{\tilde{\tau}}=150\ {\rm GeV}\ {\rm and}\
 m_{\rm LSP}=75\ {\rm GeV}, \label{masses}
\end{align}
and fix the photon energy at 
\begin{align}
 E_{\gamma}=40\ {\rm GeV}.
\label{Egamma}
\end{align}
For the $\tilde G$ LSP scenario (a),
we take four neutralino masses as 
$m_{\tilde{\chi}_{1,2,3,4}^0}=(200, 250, 300, 350)$ GeV as an example. 
The degree of polarization $P=\sqrt{P_1^2+P_2^2+P_3^2}$ is also shown with
a thick line.
Radiated photons are almost fully polarized ($P\sim1$) for the both LSP
scenarios, except around $\cos\theta=-0.95$ for the $\tilde{G}$ LSP
scenario, where photons are close to being unpolarized ($P\sim 0$).

In the $\cos\theta>0$ region, the photon bremsstrahlung amplitude
(graph 2 in Figs.~\ref{diagrams} and \ref{diagrams_n}) 
is dominant and the $\tilde G$-LSP and 
$\tilde{\chi}_1^0$-LSP cases are very similar since only
$\pm 1/2$-helicity states of the gravitino are allowed.
In the $\cos\theta<0$ region, on the other hand, 
the neutralino propagating amplitudes and the four-point interaction 
amplitude, graph 3 to 7 in
Fig.~\ref{diagrams}, become important, which allow the gravitino to take 
$\pm 3/2$ helicities as well. 
Note that the amplitude corresponding
to the graph 1 in Figs. 1 and 2 always vanishes. Since the gravitino has the large mass in
this example, spin-3/2 components dominate spin-1/2
ones, and $P_3$ for the $\tilde G$ LSP shows distinct behavior from those for 
the $\tilde{\chi}_1^0$ LSP.
Especially, for $\cos\theta\sim-1$, the difference is significant;
 $P_3=-0.8$ (almost left-handed photon) for the 
$\tilde G$ LSP, while $P_3=+1$ (right-handed photon) for the
$\tilde{\chi}_1^0$ LSP.
Those behavior holds for heavier neutralinos and agrees with the results of 
Ref.~\cite{Buchmuller:2004rq}, where the neutralino intermediate diagrams
are neglected.

Since the photon helicity measurements require a polarized detector,
we also examine linear polarizations $P_1$ and $P_2$. 
In both scenarios, the linear polarization perpendicular to the decay plane
vanishes ($P_2=0$), and $P_1$ tends to behave similarly, but
 slightly larger $|P_1|$ is expected  in
the backward direction ($\cos\theta<0$) for the 
gravitino LSP case (a).

\subsection{Neutralino NLSP}

\begin{figure}[t]
 \centering 
 \epsfig{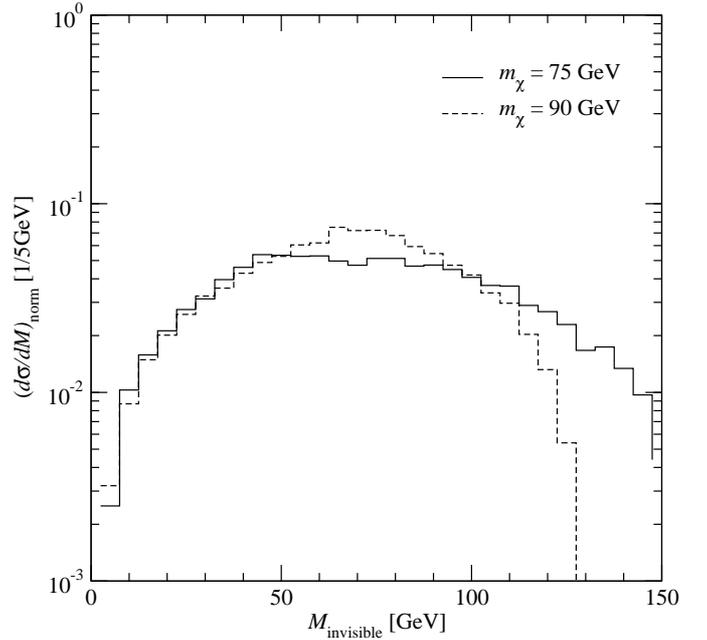}
 \caption{Missing invariant mass distributions for 
  $e^+e^-\to\tilde\chi^0_1\tilde\chi^0_1\to\gamma\gamma\tilde G\tilde G$
  at $\sqrt{s}=190$ GeV.
  The cases for the neutralino mass $m_{\chi}=75$ and 90 GeV are shown
  as a solid and dashed line, respectively, with the normalized cross section
  after kinematical cuts of \eqref{kinecut}.}
 \label{fig:minv}
\end{figure}

As a sample result for the neutralino NLSP scenario, 
we consider the process
\begin{align}
 e^+e^-\to\tilde\chi^0_1\tilde\chi^0_1
       \to(\gamma\tilde G)(\gamma\tilde G)
       \to\gamma\gamma \!\not\!\!E.
\end{align}
Figure~\ref{fig:minv} shows the distributions of the missing invariant mass 
at $\sqrt{s}=190$ GeV for the neutralino mass $m_{\chi}=75$ and 90 GeV
with the normalized cross section after kinematical cuts.
The gravitino mass is fixed at an eV order so that $\tilde\chi^0_1$
decays instantly without leaving the production point.
Here we use the same cuts as in Ref.~\cite{Ambrosanio:1996jn};
\begin{subequations}
\begin{align}
 |\cos\theta_\gamma|<0.95,\quad 
 {p_T}_\gamma>0.065\,E_{\rm beam}, \\
 0.2<E_\gamma/E_{\rm beam}<0.8,
\end{align}
\label{kinecut}%
\end{subequations}
with $E_{\rm beam}=\sqrt{s}/2$,
and our results agree well with Fig.~16 in~\cite{Ambrosanio:1996jn}.

%%%%%%%%%%%%%% Begin Section 3 %%%%%%%%%%%%%%%%%%%%%%%%%%%%%%%%%%%%%%%%% 
\section{Summary}\label{sec:summary}

In this paper, we have added new {\tt HELAS} 
subroutines to calculate helicity amplitudes with
massive spin-3/2 particles (massive gravitinos) 
to the {\tt HELAS} library.
They are
coded in such a way that arbitrary amplitudes with external gravitinos 
can be generated automatically by {\tt MG}, after slight
modifications. All the codes have been tested carefully by making use of 
the gauge invariance of the helicity amplitudes.

%AAAAAAAAAAAAAAAAAAAAAAAAAAAAAAAAAAAAAAAAAAAAAAAAA
\begin{acknowledgement}{\textit{Acknowledgements}}
We wish to thank Qiang Li for helping us modify {\tt MadGraph}
and Junichi Kanzaki for putting our code on the web.
K.H. and Y.T. would like to thank Tilman Plehn and the members of 
the ITP, Uni.~Heidelberg for their warm hospitality, 
where part of this work has been done.
The work presented here has been in part supported by the Concerted
 Research action 
``Supersymmetric Models and their Signatures at the Large Hadron
 Collider'' 
of the Vrije Universiteit Brussel,
by the IISN ``MadGraph'' convention 4.4511.10,
by the Belgian Federal Science Policy Office through the Interuniversity
 Attraction Pole IAP VI/11, 
and by the Grant-in-Aid for Scientific Research (No. 20340064) from the
 Japan Society for the Promotion of Science.  
Y.T. was also supported in part by Institutional Program for Young
 Researcher Overseas Visits.
\end{acknowledgement}

\appendix
%%%%%%%%%%%%%% Begin App. 1 %%%%%%%%%%%%%%%%%%%%%%%%%%%%%%%%%%%%%%%%
\section{HELAS subroutines for spin-3/2 particles}\label{sec:helas_new}

In this appendix, we list the contents of all the new {\tt HELAS}
subroutines that are needed to evaluate processes based on the 
effective Lagrangian of \eqref{L_int} with external 
spin-3/2 gravitinos.
 
To begin with,  in App.~\ref{sec:wavefunc} the subroutines to compute
external lines for a massive spin-3/2 particle are presented.
Next, in Apps.~\ref{sec:vertex_i} to
\ref{sec:vertex_f}, we explain vertex subroutines listed in
Table~\ref{sublist_new}, which compute interactions of a
gravitino with SM and SUSY particles.
Finally, we briefly mention how we test our new subroutines in
App.~\ref{sec:check}.

%%%%%%%%%%%%%%%%%%%%%%%%%%%%%%%%%%%%%%%%%%%%%%%%%%%%%%%
\subsection{Spin-3/2 wavefunction}\label{sec:wavefunc}

\subsubsection{\tt IRXXXX}\label{sec:irxxxx}

This subroutine computes the flowing-{\tt I}n {\tt R}arita-Schwinger
({\tt R}-S) spin-3/2 wavefunction; namely $\psi^{\mu}_u(p,\lambda)$
and $\psi^{\mu}_v(p,\lambda)$, in terms of its four-momentum $p$
and helicity $\lambda$, and should be called as
\begin{align*}
  {\tt CALL\ IRXXXX(P,RMASS,NHEL,NSR\ ,\ RI)}.
\end{align*}
The input {\tt P(0:3)} is a real four-dimensional array which
contains the four-momentum $p^{\mu}$ of the spin-3/2 particle,
{\tt RMASS} is its mass, {\tt NHEL}
(${\tt =\pm 3,\pm 1}$) specifies its helicity $\lambda$ in unit of 1/2,
and {\tt NSR} 
specifies whether the fermion is particle or anti-particle.
If {\tt NSR = 1} the fermion is particle and the subroutine computes the
wavefunction with the $u$-spinor. 
If {\tt NSR = -1} the fermion is anti-particle and the subroutine
computes the wavefunction with the $v$-spinor.%
\footnote{Although the gravitino is a Majorana particle, the {\tt HELAS}
convention requires both of the $u$- and $v$-spinors for the
calculations of amplitudes and their proper interference; 
see App.~A in~\cite{Cho:2006sx}.} 
The output {\tt RI(18)} is a complex
18-dimensional array, among which the first 16 components contain the
wavefunction as
\begin{align}
 {\tt RI(4\mu+i)=R(\mu+1,i)},
\label{TtoRI}
\end{align}
namely
\begin{align*}
 &{\tt RI(\ 1)=R(1,1)}, \nn \\
 &{\tt RI(\ 2)=R(1,2)}, \nn \\
 &{\tt RI(\ 3)=R(1,3)}, \nn \\
 &{\tt RI(\ 4)=R(1,4)}, \nn \\
 &{\tt ......       } \nn \\
 &{\tt RI(16)=R(4,4)},
\end{align*}
where
\begin{align}
 {\tt R(\mu+1,i)}=
 \begin{cases}
  \psi^{\mu}_{u_i}(p,\lambda) &\text{for {\tt NSR = 1}}, \\
  \psi^{\mu}_{v_i}(p,\lambda) &\text{for {\tt NSR = -1}}.
 \end{cases}
\label{Tmunu}
\end{align}
Here, $i=1,2,3,4$ denotes each $u$- or $v$-spinor component.
The last two of {\tt RI(18)} contain the four-momentum along the fermion
number flow,
\begin{align}
  ({\tt RI(17)},\,{\tt RI(18)})
 ={\tt NSR}\,({\tt P(0)}+i{\tt P(3)},\,{\tt P(1)}+i{\tt P(2)}).
\end{align}

\begin{table}
\centering
\begin{tabular}{|c|c|c|c|} \hline
 Vertex & Inputs & Output & Subroutine \\ \hline\hline
 FRS & FRS & Amplitude & {\tt IORSXX}, {\tt IROSXX} \\
     & RS  & F         & {\tt FSORXX}, {\tt FSIRXX} \\ 
     & FR  & S         & {\tt HIORXX}, {\tt HIROXX} \\ \hline
 FRV & FRV & Amplitude & {\tt IORVXX}, {\tt IROVXX} \\
     & RV  & F         & {\tt FVORXX}, {\tt FVIRXX} \\
     & FR  & V         & {\tt JIORXX}, {\tt JIROXX} \\ \hline
 FRVS & FRVS & Amplitude & {\tt IORVSX}, {\tt IROVSX} \\
      & RVS  & F         & {\tt FVSORX}, {\tt FVSIRX} \\
      & FRS  & V         & {\tt JSIORX}, {\tt JSIROX} \\
      & FRV  & S         & {\tt HVIORX}, {\tt HVIROX} \\ \hline
 FRVV & FRVV & Amplitude & {\tt IORVVX}, {\tt IROVVX} \\
      & RVV  & F         & {\tt FVVORX}, {\tt FVVIRX} \\
      & FRV  & V         & {\tt JVIORX}, {\tt JVIROX} \\ \hline
\end{tabular}
\caption{List of the new vertex subroutines in {\tt HELAS} system.}
\label{sublist_new}
\end{table}

When the four-momentum of the R-S fermion is given by
\begin{align}
 p^{\mu}=(E,\,|\vec p|\sin\theta\cos\phi,\,
      |\vec p|\sin\theta\sin\phi,\,|\vec p|\cos\theta),
\end{align}
its helicity states can be expressed as
\begin{align}
 \psi^{\mu}_{u}(p,+3/2) &=
   \epsilon^\mu(p,+)\,u(p,+), \nn \\
 \psi^{\mu}_{u}(p,+1/2) &=
   \sqrt{\frac{2}{3}}\,\epsilon^\mu(p,0)\,u(p,+) \nn\\
 &\quad+\sqrt{\frac{1}{3}}\,\epsilon^\mu(p,+)\,u(p,-)\,e^{i\phi}, \nn\\
 \psi^{\mu}_{u}(p,-1/2) &=
   \sqrt{\frac{1}{3}}\,\epsilon^\mu(p,-)\,u(p,+) \nn\\
 &\quad+\sqrt{\frac{2}{3}}\,\epsilon^\mu(p,0)\,u(p,-)\,e^{i\phi}, \nn\\
 \psi^{\mu}_{u}(p,-3/2) &=
   \epsilon^\mu(p,-)\,u(p,-)\,e^{i\phi},
\end{align}
by using the vector boson wavefunctions $\epsilon^\mu(p,\lambda)$ and
the spinor wavefunctions $u(p,\lambda)$ that obey the relations
\begin{align}
 J_-\,\eps^{\mu}(p,\lambda)&=\sqrt{2}\,\eps^{\mu}(p,\lambda-1), \\
 J_-\,u(p,+)&=e^{i\phi}\,u(p,-),
\end{align}
where $J_-=J_x-iJ_y$ is the $J_z$ lowering operator. The vector and spinor
wavefunctions in the {\tt HELAS} convention~\cite{Hagiwara:1990dw}
satisfy above relations.
Similarly, $\psi^{\mu}_{v}(p,\lambda)$ is given by the $v$-spinors and the
conjugated vector wavefunctions. 
The above helicity states satisfy the irreducibility conditions and the
Dirac equation, 
\begin{align}
 \gamma_{\mu}\psi^{\mu}_{u}(p,\lambda) =0,\quad 
 p_{\mu}\psi^{\mu}_{u}(p,\lambda) =0, \\
 (\not\!p-m_{3/2})\psi^{\mu}_{u}(p,\lambda) =0,
\end{align}
and the completeness relation is
\begin{align}
 \sum_{\lambda=-3/2}^{+3/2}
  \psi^{\mu}_{u}(p,\lambda)\overline\psi^{\nu}_{u}(p,\lambda)
 =P^{\mu\nu}(p),
\end{align}
where
\begin{multline}
 P^{\mu\nu}(p)= (\not\!p+m_{3/2})
  \Big(
   \Pi^{\mu\nu}(p)
   +\frac{1}{3}
    \Pi^{\mu\alpha}(p)\Pi^{\nu\beta}(p)
    \gamma_{\alpha}\gamma_{\beta}
  \Big)
\end{multline}
with
\begin{align}
 \Pi^{\mu\nu}(p)=-g^{\mu\nu}+\frac{p^{\mu}p^{\nu}}{m_{3/2}^2}.
\end{align}

\subsubsection{\tt ORXXXX}\label{sec:orxxxx}

This subroutine computes the flowing-{\tt O}ut {\tt R}-S
wavefunction; namely, $\overline\psi^{\mu}_u(p,\lambda)$ and 
$\overline\psi^{\mu}_v(p,\lambda)$,
and should be called as
\begin{align*}
  {\tt CALL\ ORXXXX(P,RMASS,NHEL,NSR\ ,\ RO)}.
\end{align*}
As in the subroutine {\tt IRXXXX}, the output {\tt RO(18)} is a complex
18-dimensional array, among which the first 16 components contain the
wavefunction as
\begin{align}
 {\tt RO(4\mu+i)=\bar{R}(\mu+1,i)},
\label{RtoRO}
\end{align}
where
\begin{align}
 {\tt \bar{R}(\mu+1,i)}=
 \begin{cases}
  \overline\psi^{\mu}_{u_i}(p,\lambda) &\text{for {\tt NSR = 1}}, \\
  \overline\psi^{\mu}_{v_i}(p,\lambda) &\text{for {\tt NSR = -1}},
 \end{cases}
\end{align}
and the last two are the four-momentum
\begin{align}
  ({\tt RO(17)},\,{\tt RO(18)})
 ={\tt NSR}\,({\tt P(0)}+i{\tt P(3)},\,{\tt P(1)}+i{\tt P(2)}).
\end{align}
%

%%%%%%%%%%%%%%%%%%%%%%%%%%%%%%%%%%%%%%%%%%%%%%%%%%%%%%%%%%%%%%%%%%
\subsection{FRS vertex}\label{sec:vertex_i}

The {\tt FRS} vertices are obtained from the interaction Lagrangian
among a fermion, a R-S fermion and a scalar boson:
\begin{align}
 {\cal L}_{\tt FRS}=-i\overline{R}_{\mu}\gamma^{\nu}\gamma^{\mu}
  [{\tt GR(1)}P_L+{\tt GR(2)}P_R]f\,\partial_{\nu}S^*+{\rm h.c.}
\label{L_FRS}
\end{align}
with the notation $R^{\mu}=\psi^{\mu}_{u/v}$ and 
the chiral-projection operator $P_{R/L}=\frac{1}{2}(1\pm\gamma_5)$.
{\tt GR(1)} and {\tt GR(2)} are the relevant left and right coupling
constants.
For instance, in the case of the quark-gravitino-squark
interaction, $q$-$\tilde G$-$\tilde q_{\alpha}$, those couplings
are 
\begin{align}
 &\begin{cases}
  {\tt GR(1)}={\tt GFRSL(1)}={\tt GFRS}\\ 
  {\tt GR(2)}={\tt GFRSL(2)}=0
 \end{cases} &&{\rm for}\ \alpha=L, \label{GFRSL}\\
 &\begin{cases}
  {\tt GR(1)}={\tt GFRSR(1)}=0\\ 
  {\tt GR(2)}={\tt GFRSR(2)}=-{\tt GFRS}
 \end{cases} &&{\rm for}\ \alpha=R, \label{GFRSR}
\end{align}
where
\begin{align}
 {\tt GFRS} = 1/\sqrt{2}\,\Mpl.
\label{GFRS}
\end{align}

\subsubsection{\tt IORSXX}

This subroutine computes an amplitude of the {\tt FRS} vertex from
wavefunctions of a
flowing-{\tt I}n fermion, a flowing-{\tt O}ut {\tt R}-S 
fermion and a {\tt S}calar boson, and should
be called as
\begin{center}
 {\tt CALL\ IORSXX(FI,RO,SC,GR , VERTEX)}.
\end{center}
The input {\tt FI(6)} is a complex six-dimensional array which contains the
wavefunction of the flowing-{\tt I}n {\tt F}ermion and its four-momentum as 
\begin{align*} 
 p^{\mu}=(\Re e{\tt FI(5)},\Re e{\tt FI(6)},
          \Im m{\tt FI(6)},\Im m{\tt FI(5)}).
\end{align*}
The input {\tt RO(18)} is a complex 18-dimensional array which consists of the
wavefunction and the four-momentum of the flowing-{\tt O}ut {\tt R}-S
fermion; see 
the {\tt ORXXXX} subroutine in App.~\ref{sec:orxxxx}, 
while the input {\tt SC(3)} is a complex
three-dimensional array which contains the wavefunction of the 
{\tt S}calar boson, {\tt SC(1)}, and its four-momentum as
\begin{align*}
 q^{\mu}=(\Re e{\tt SC(2)},\Re e{\tt SC(3)},
          \Im m{\tt SC(3)},\Im m{\tt SC(2)}).
\end{align*}
The input {\tt GR(2)} is the complex coupling constant, such as in
(\ref{GFRSL}) and (\ref{GFRSR}) in units of GeV$^{-1}$. 
The output {\tt VERTEX} is a complex number in units of GeV:
\begin{equation}
 {\tt VERTEX}=({\tt RO})_{\mu}{\tt SC(1)}\qs\gamma^{\mu}
  [{\tt GR(1)}P_{L}+{\tt GR(2)}P_{R}]({\tt FI}),
\label{iors}
\end{equation}
where we use the notations
\begin{align}
 ({\tt FI})&=\begin{pmatrix}
	      {\tt FI(1)}\\{\tt FI(2)}\\{\tt FI(3)}\\{\tt FI(4)}
	     \end{pmatrix},\\
 ({\tt RO})_{\mu}&= 
  {\tt (RO(4\mu+1),RO(4\mu+2),RO(4\mu+3),RO(4\mu+4))}.
\end{align}

\subsubsection{\tt IROSXX}

This subroutine computes an amplitude of the {\tt FRS} vertex from
wavefunctions of
a flowing-{\tt I}n {\tt R}-S fermion,
a flowing-{\tt O}ut fermion and a {\tt S}calar
boson, and should be called as
\begin{center}
 {\tt CALL IROSXX(RI,FO,SC,GR , VERTEX)}.
\end{center}
The input {\tt RI(18)} is a complex 18-dimensional array which consists
of the wavefunction and the four-momentum of the flowing-{\tt I}n 
{\tt R}-S fermion; see
the {\tt IRXXXX} subroutine in App.~\ref{sec:irxxxx},
while the input {\tt FO(6)} is a complex
six-dimensional array which contains the
wavefunction of the flowing-{\tt O}ut {\tt F}ermion and its four-momentum as
\begin{align*}
 p^{\mu}=(\Re e{\tt FO(5)},\Re e{\tt FO(6)},
          \Im m{\tt FO(6)},\Im m{\tt FO(5)}).
\end{align*}
The output {\tt VERTEX} is a complex number:
\begin{equation}
 {\tt VERTEX}=-{\tt (FO)}{\tt SC(1)}
  [{\tt GR(1)}^*P_{R}+{\tt GR(2)^*}P_{L}]\gamma^{\mu}\qs
 ({\tt RI})_{\mu},
\end{equation}
where $q^{\mu}$ is the momentum of the scalar boson 
and we use the notations
\begin{align}
 ({\tt RI})_{\mu}&=\begin{pmatrix}
		    {\tt RI(4\mu+1)}\\{\tt RI(4\mu+2)}\\
		    {\tt RI(4\mu+3)}\\{\tt RI(4\mu+4)}
	           \end{pmatrix},\\
 {\tt (FO)}&={\tt (FO(1),FO(2),FO(3),FO(4)}).
\end{align}

\subsubsection{\tt FSORXX}

This subroutine computes an off-shell {\tt F}ermion wavefunction made
from the interaction of a {\tt S}calar boson and a flowing-{\tt O}ut
{\tt R}-S fermion by the {\tt FRS} vertex, and should be called as
\begin{center}
 {\tt CALL FSORXX(RO,SC,GR,FMASS,FWIDTH , FSOR)},
\end{center}
where {\tt FMASS} and {\tt FWIDTH} are the mass and the width of the fermion,
$m_F$ and $\Gamma_F$.
 The output {\tt FSOR(6)} gives the off-shell fermion wavefunction
multiplied by the fermion propagator and its four-momentum, which is
expressed as a complex six-dimensional array: 
\begin{multline}
 {\tt (FSOR)}=({\tt RO})_{\mu}{\tt SC(1)}\qs\gamma^{\mu}
  [i{\tt GR(1)}P_{L}+i{\tt GR(2)}P_{R}] \\
 \times\frac{i(\not\!p+m_{F})}{p^2-m_{F}^2+im_{F}\Gamma_{F}},
\label{fsor}
\end{multline}
and
\begin{align}
 {\tt FSOR(5)}&={\tt RO(17)+SC(2)},\label{fsor5}\\
 {\tt FSOR(6)}&={\tt RO(18)+SC(3)}.\label{fsor6}
\end{align}
Here we use the notation
\begin{align}
 ({\tt FSOR})=({\tt FSOR(1),FSOR(2),FSOR(3),FSOR(4)}),
\end{align}
and $p$ is the momentum of the off-shell fermion given in (\ref{fsor5})
and (\ref{fsor6}) as 
\begin{align*}
 p^{\mu}=(\Re e{\tt FSOR(5)},\Re e{\tt FSOR(6)},
          \Im m{\tt FSOR(6)},\Im m{\tt FSOR(5)}).
\end{align*}

\subsubsection{\tt FSIRXX}

The subroutine computes an off-shell {\tt F}ermion wavefunction 
made from the interaction of a {\tt S}calar boson and
a flowing-{\tt I}n {\tt R}-S fermion by the {\tt FRS} vertex, 
and should be called as
\begin{center}
 {\tt CALL FSIRXX(RI,SC,GR,FMASS,FWIDTH , FSIR)}.
\end{center}
The output {\tt FSIR(6)} 
is a complex six-dimensional array:
\begin{multline}
{\tt (FSIR)}
 =-\frac{i(\not\!p+m_{F})}{p^2-m_{F}^2+im_{F}\Gamma_{F}}\,{\tt SC(1)}\\
   \times[i{\tt
 GR(1)}^*P_{R}+i{\tt GR(2)}^*P_{L}]
 \gamma^{\mu}\qs\,({\tt RI})_{\mu},
\end{multline}
and
\begin{align}
 {\tt FSIR(5)}&={\tt RI(17)-SC(2)},\\
 {\tt FSIR(6)}&={\tt RI(18)-SC(3)}.
\end{align}
Here we use the notation
\begin{equation}
 ({\tt FSIR})=\begin{pmatrix}
	       {\tt FSIR(1)}\\
	       {\tt FSIR(2)}\\
	       {\tt FSIR(3)}\\
	       {\tt FSIR(4)}
	      \end{pmatrix},
\end{equation}
and the momentum $p$ is
\begin{align*}
 p^{\mu}=(\Re e{\tt FSIR(5)},\Re e{\tt FSIR(6)},
          \Im m{\tt FSIR(6)},\Im m{\tt FSIR(5)}).
\end{align*}

\subsubsection{\tt HIORXX}

This subroutine computes an off-shell scalar current {\tt H} made from
the interaction of a flowing-{\tt I}n fermion and  a flowing-{\tt O}ut 
{\tt R}-S fermion by the
{\tt FRS} vertex, and should be called as
\begin{center}
 {\tt CALL HIORXX(FI,RO,GR,SMASS,SWIDTH , HIOR)},
\end{center}
where {\tt SMASS} and {\tt SWIDTH} are the mass and the width of the
scalar boson, $m_S$ and $\Gamma_S$.
The output {\tt HIOR(3)} gives the off-shell scalar current
multiplied by the scalar boson propagator and its four-momentum, which
is expressed as a complex three-dimensional array:
\begin{multline}
 {\tt HIOR(1)} =-\frac{i}{q^2-m_{S}^2+im_{S}\Gamma_{S}}\\
 \times({\tt RO})_{\mu}\qs\gamma^{\mu}
  [i{\tt GR(1)}P_{L}+i{\tt GR(2)}P_{R}]({\tt FI}),
\end{multline}
and
\begin{align}
 {\tt HIOR(2)}&={\tt -FI(5)+RO(17)},\\
 {\tt HIOR(3)}&={\tt -FI(6)+RO(18)}.
\end{align}
The momentum $q$ is
\begin{align*}
 q^{\mu}=(\Re e{\tt HIOR(2)},\Re e{\tt HIOR(3)},
          \Im m{\tt HIOR(3)},\Im m{\tt HIOR(2)}).
\end{align*}

\subsubsection{\tt HIROXX}

This subroutine computes an off-shell scalar current {\tt H} made from
the interaction of a flowing-{\tt I}n {\tt R}-S fermion and a
flowing-{\tt O}ut fermion by the {\tt FRS} vertex, and should be called as
\begin{center}
 {\tt CALL HIROXX(RI,FO,GR,SMASS,SWIDTH , HIRO)}.
\end{center}
The output {\tt HIRO(3)} 
is a complex three-dimensional array:
\begin{multline}
 {\tt HIRO(1)}=\frac{i}{q^2-m_{S}^2+im_{S}\Gamma_{S}} \\
 \times({\tt FO})[i{\tt GR(1)}^*P_{R}+i{\tt GR(2)}^*P_{L}]
  \gamma^{\mu}\qs({\tt RI})_{\mu},
\end{multline}
and
\begin{align}
 {\tt HIRO(2)}&={\tt -RI(17)+FO(5)},\\
 {\tt HIRO(3)}&={\tt -RI(18)+FO(6)}.
\end{align}
The momentum $q$ is
\begin{align*}
 q^{\mu}=(\Re e{\tt HIRO(2)},\Re e{\tt HIRO(3)},
          \Im m{\tt HIRO(3)},\Im m{\tt HIRO(2)}).
\end{align*}

Before turning to the {\tt FRV} vertex, it should be noticed here that
the conventional factors of $i$ in the vertices and those in the
propagators are both included in the off-shell wavefunctions, such as
(\ref{fsor}) above, according to the {\tt HELAS} convention. The
{\tt HELAS} amplitude, obtained by the vertices, such as
(\ref{iors}), gives the contribution to the $T$ matrix element
without the factor of $i$. See more details in the {\tt HELAS} 
manual~\cite{Hagiwara:1990dw}.

%%%%%%%%%%%%%%%%%%%%%%%%%%%%%%%%%%%%%%%%%%%%%%%%%%%%%%%%%%%%%%%%%%%%%%%%
\subsection{FRV vertex}

The {\tt FRV} vertices are obtained from the interaction Lagrangian
among a fermion, a R-S fermion and a vector boson:
\begin{equation}
 {\cal L}_{\tt FRV}=-i\overline{R}_{\mu}^{}[\gamma^{\nu},\gamma^{\rho}]
  \gamma^{\mu}[{\tt GR(1)}P_L+{\tt GR(2)}P_R]f\,
  \partial_{\nu}^{}V^*_{\rho}
 +{\rm h.c.}
\end{equation}
We note that, although both a gravitino and a gaugino are
Majorana in most cases, the Hermitian conjugate term is necessary for
{\tt MG}; practically, either the first or second term is used in
calculations of amplitudes. 
The corresponding coupling constant to the effective Lagrangian 
of (\ref{L_int}) is 
\begin{align}
 {\tt GR(1)}={\tt GR(2)}={\tt GFRV}=1/4\Mpl.
\label{GFRV}
\end{align}

\subsubsection{\tt IORVXX}

This subroutine computes an amplitude of the {\tt FRV} vertex from
wavefunctions of a flowing-{\tt I}n fermion, a flowing-{\tt O}ut 
{\tt R}-S fermion and a {\tt V}ector boson, and should
be called as
\begin{center}
 {\tt CALL IORVXX(FI,RO,VC,GR , VERTEX)}.
\end{center}
The input {\tt VC(6)} is a complex six-dimensional array which contains
the {\tt V}ector boson wavefunction and its momentum as
\begin{align*}
 q^{\mu}=(\Re e{\tt VC(5)},\Re e{\tt VC(6)},
          \Im m{\tt VC(6)},\Im m{\tt VC(5)}).
\end{align*}
The input {\tt GR} is the coupling constant in (\ref{GFRV}). The
output {\tt VERTEX} is a complex number:
\begin{align}
 {\tt VERTEX}=({\tt RO})_{\mu}[\not\!q,\not\!V]\gamma^{\mu}
  [{\tt GR(1)}P_{L}+{\tt GR(2)}P_{R}]({\tt FI}), 
\end{align}
where we use the notation
\begin{align}
 V^{\mu}={\tt VC}(\mu+1).
\end{align}

\subsubsection{\tt IROVXX}

This subroutine computes an amplitude of the {\tt FRV} vertex from
wavefunctions of a flowing-{\tt I}n {\tt R}-S fermion, a 
flowing-{\tt O}ut fermion and a {\tt V}ector boson, and should be 
called as
\begin{center}
 {\tt CALL IROVXX(RI,FO,VC,GR , VERTEX)}.
\end{center}
The output {\tt VERTEX} is
\begin{align}
 {\tt VERTEX}=-({\tt FO})
  [{\tt GR(1)}^*P_{R}+{\tt GR(2)}^*P_{L}]\gamma^{\mu}
  [\not\!V,\not\!q]({\tt RI})_{\mu}.
\end{align}

\subsubsection{\tt FVORXX}

This subroutine computes an off-shell {\tt F}ermion wavefunction made
from the interaction of a {\tt V}ector boson and a flowing-{\tt O}ut 
{\tt R}-S fermion by the {\tt FRV} vertex, and should be called as
\begin{center}
 {\tt CALL FVORXX(RO,VC,GR,FMASS,FWIDTH , FVOR)}.
\end{center}
What we compute here is
\begin{multline}
 {\tt (FVOR)}=({\tt RO})_{\mu}[\not\!q,\not\!V]\gamma^{\mu}
  [i{\tt GR(1)}P_{L}+i{\tt GR(2)}P_{R}] \\
 \times\frac{i(\not\!p+m_{F})}{p^2-m_{F}^2+im_{F}\Gamma_{F}},
\end{multline}
and
\begin{align}
 {\tt FVOR(5)}&={\tt RO(17)+VC(5)}, \\
 {\tt FVOR(6)}&={\tt RO(18)+VC(6)},
\end{align}
where we use the notation
\begin{align}
 ({\tt FVOR})=&({\tt FVOR(1),FVOR(2),FVOR(3),FVOR(4)}),
\end{align}
and the momentum $p$ is
\begin{align*}
 p^{\mu}=(\Re e{\tt FVOR(5)},\Re e{\tt FVOR(6)},
          \Im m{\tt FVOR(6)},\Im m{\tt FVOR(5)}).
\end{align*}

\subsubsection{\tt FVIRXX}

This subroutine computes an off-shell {\tt F}ermion wavefunction made
from the interaction of a {\tt V}ector boson and a flowing-{\tt I}n
{\tt R}-S fermion by the {\tt FRV} vertex, and should be called as
\begin{center}
 {\tt CALL FVIRXX(RI,VC,GR,FMASS,FWIDTH , FVIR)}.
\end{center}
What we compute here is
\begin{multline}
 {\tt (FVIR)}=-\frac{i(\not\!p+m_{F})}{p^2-m_{F}^2+im_{F}\Gamma_{F}}\\
 \times [i{\tt GR(1)}^*P_{R}+i{\tt GR(2)}^*P_{L}]
   \gamma^{\mu}[\not\!V,\not\!q]({\tt RI})_{\mu},
\end{multline}
and
\begin{align}
 {\tt FVIR(5)}&={\tt RI(17)-VC(5)},\\
 {\tt FVIR(6)}&={\tt RI(18)-VC(6)},
\end{align}
where we use the notation
\begin{align}
 ({\tt FVIR})=&\begin{pmatrix}
	      {\tt FVIR(1)}\\
	      {\tt FVIR(2)}\\
	      {\tt FVIR(3)}\\
	      {\tt FVIR(4)}
	     \end{pmatrix},
\end{align}
and the momentum $p$ is
\begin{align*}
 p^{\mu}=(\Re e{\tt FVIR(5)},\Re e{\tt FVIR(6)},
          \Im m{\tt FVIR(6)},\Im m{\tt FVIR(5)}).
\end{align*}

\subsubsection{\tt JIORXX}

This subroutine computes an off-shell vector current {\tt J} made from
the interaction of a flowing-{\tt I}n fermion and a flowing-{\tt O}ut
{\tt R}-S fermion by the {\tt FRV} vertex, and should be called as
\begin{center}
 {\tt CALL JIORXX(FI,RO,GR,VMASS,VWIDTH , JIOR)}.
\end{center}
The input {\tt VMASS} and {\tt VWIDTH} are the mass and the width of the
vector boson, $m_V$ and $\Gamma_V$. The output {\tt JIOR(6)} gives the off-shell
vector current multiplied by the vector boson propagator and its
four-momentum, which is expressed as a complex six-dimensional array:
\begin{multline}
 {\tt JIOR(\nu+1)}=-\frac{i}{q^2-m_{V}^2+im_{V}\Gamma_{V}}
  \left(-g^{\rho\nu}+\frac{q^{\rho}q^{\nu}}{m_{V}^{2}}\right) \\
  \times ({\tt RO})_{\mu}
  [\not\!q,\gamma_{\rho}]\gamma^{\mu}
  [i{\tt GR(1)}P_{L}+i{\tt GR(2)}P_{R}]({\tt FI})
\end{multline}
for the massive vector boson, or
\begin{multline}
 {\tt JIOR(\nu+1)}=-\frac{-i}{q^2} \\
  \times({\tt RO})_{\mu}[\not\!q,\gamma^{\nu}]\gamma^{\mu}
 [i{\tt GR(1)}P_{L}+i{\tt GR(2)}P_{R}]({\tt FI})
\end{multline}
for the massless vector boson, and 
\begin{align}
 {\tt JIOR(5)}&={\tt -FI(5)+RO(17)},\\
 {\tt JIOR(6)}&={\tt -FI(6)+RO(18)}.
\end{align}
Here, $q$ is the momentum of the off-shell vector boson,
\begin{align*}
  q^{\mu}=&({\tt \Re eJIOR(5),\Re eJIOR(6),\Im mJIOR(6),\Im mJIOR(5)}).
\end{align*}
Note that we use the unitary gauge for the massive vector boson
propagator and the Feynman gauge for the massless one, according to
the {\tt HELAS} convention~\cite{Hagiwara:1990dw}.

\subsubsection{\tt JIROXX}

This subroutine computes an off-shell vector current {\tt J} made from
the interaction of a flowing-{\tt I}n {\tt R}-S fermion and a
flowing-{\tt O}ut fermion by the {\tt FRV} vertex, and should be called
as
\begin{center}
 {\tt CALL JIROXX(RI,FO,GR,VMASS,VWIDTH , JIRO)}.
\end{center}
The output {\tt JIRO(6)} is
\begin{multline}
 {\tt JIRO(\nu+1)}=\frac{i}{q^2-m_{V}^{2}+im_{V}\Gamma_{V}}
  \left(-g^{\rho\nu}+\frac{q^{\rho}q^{\nu}}{m_{V}^{2}}\right) \\
  \times({\tt FO})
  [i{\tt GR(1)}^*P_{R}+i{\tt GR(2)}^*P_{L}]\gamma^{\mu}
  [\gamma_{\rho},\not\!q]({\tt RI})_{\mu}
\end{multline}
for the massive vector boson, or
\begin{multline}
 {\tt JIRO(\nu+1)}=\frac{-i}{q^2} \\
 \times({\tt FO})[i{\tt GR(1)}^*P_{R}+i{\tt GR(2)}^*P_{L}]
 \gamma^{\mu}[\gamma^{\nu},\not\!q]({\tt RI})_{\mu}
\end{multline}
for the massless vector boson, and 
\begin{align}
 {\tt JIRO(5)}&={\tt -RI(17)+FO(5)},\\
 {\tt JIRO(6)}&={\tt -RI(18)+FO(6)}.
\end{align}
Here the momentum $q$ is
\begin{align*}
  q^{\mu}=({\tt \Re eJIRO(5),\Re eJIRO(6),\Im mJIRO(6),\Im mJIRO(5)}).
\end{align*}
%

%%%%%%%%%%%%%%%%%%%%%%%%%%%%%%%%%%%%%%%%%%%%%%%%%%%%%%%%%%%%%%%%%%%%
\subsection{FRVS vertex}

The {\tt FRVS} vertices are obtained from the interaction Lagrangian
among a fermion, a R-S fermion, a vector boson and a scalar boson: 
\begin{align}
 {\cal L}_{\tt FRVS}=
 \overline{R}_{\mu}^{}\gamma^{\nu}\gamma^{\mu}
 [{\tt GR(1)}P_L+{\tt GR(2)}P_R]f\,V_{\nu}^*S^* + {\rm h.c.}
\end{align}
The coupling constant {\tt GR} is the product of 
the {\tt FRS} coupling constant
and the gauge coupling constant of the involving gauge boson.
For instance, in the case of the quark-gravitino-gluon-squark
interaction, $q$-$\tilde G$-$g$-$\tilde q_{L}$, those couplings
are 
\begin{align}
 \begin{cases}
  {\tt GR(1)}={\tt GFRGSL(1)}={\tt GFRSL(1)*GG(1)},\\ 
  {\tt GR(2)}={\tt GFRGSL(2)}={\tt GFRSL(2)*GG(2)},
 \end{cases} 
\label{GR_GFRVS}
\end{align}
where {\tt GFRSL} is defined in (\ref{GFRSL}) and {\tt GG} is
the strong coupling constant
\begin{align}
 {\tt GG(1)} = {\tt GG(2)} = -g_s =-{\tt G}.
\label{GG}
\end{align}
The sign of the coupling constant is fixed by the {\tt HELAS}
convention~\cite{Hagiwara:1990dw}.

\subsubsection{\tt IORVSX}

This subroutine computes an amplitude of the {\tt FRVS} vertex from a
flowing-{\tt I}n fermion, a flowing-{\tt O}ut {\tt R}-S fermion, a 
{\tt V}ector boson and a {\tt S}calar boson, and should be called as
\begin{center}
 {\tt CALL IORVSX(FI,RO,VC,SC,GR , VERTEX)}.
\end{center}
The output {\tt VERTEX} gives a complex number:
\begin{align}
 {\tt VERTEX}=({\tt RO})_{\mu}{\tt SC(1)}\!\not\!V\gamma^{\mu}
  [{\tt GR(1)}P_{L}+{\tt GR(2)}P_{R}]({\tt FI}).
\end{align}

\subsubsection{\tt IROVSX}

This subroutine computes an amplitude of the {\tt FRVS} vertex from a
flowing-{\tt I}n {\tt R}-S fermion, a flowing-{\tt O}ut fermion, a
{\tt V}ector boson and a {\tt S}calar boson, and should be called as
\begin{center}
 {\tt CALL IROVSX(RI,FO,VC,SC,GR , VERTEX)}.
\end{center}
The output {\tt VERTEX} gives a complex number:
\begin{align}
 {\tt VERTEX}=({\tt FO})\,{\tt SC(1)}
  [{\tt GR(1)}^*P_{R}+{\tt GR(2)}^*P_{L}]\gamma^\mu\!\not\!V({\tt RI})_{\mu}.
\end{align}

\subsubsection{\tt FVSORX}

This subroutine computes an off-shell {\tt F}ermion wavefunction 
made from the interaction of a {\tt V}ector boson, a {\tt S}calar
boson and a flowing-{\tt O}ut {\tt R}-S fermion 
by the {\tt FRVS} vertex, and should be called as 
\begin{align*}
 {\tt CALL\ FVSORX(RO,VC,SC,GR,FMASS,FWIDTH\ ,\ FVSOR)}.
\end{align*}
The output {\tt FVSOR} is a complex six-dimensional array:
\begin{multline}
 {\tt (FVSOR)}=({\tt RO})_{\mu}{\tt SC(1)}\!\not\!V\gamma^{\mu}
  [i{\tt GR(1)}P_{L}+i{\tt GR(2)}P_{R}] \\
 \times\frac{i(\not\!p+m_{F})}{p^2-m_{F}^2+im_{F}\Gamma_{F}}
\end{multline}
for the first four components of {\tt FVSOR(6)}, and 
\begin{align}
 {\tt FVSOR(5)}&={\tt RO(17)+VC(5)+SC(2)},\\
 {\tt FVSOR(6)}&={\tt RO(18)+VC(6)+SC(3)},
\end{align}
for the momentum $p$.

\subsubsection{\tt FVSIRX}

This subroutine computes an off-shell {\tt F}ermion wavefunction 
made from the interaction of a {\tt V}ector boson, a {\tt S}calar
boson and a flowing-{\tt I}n {\tt R}-S fermion
by the {\tt FRVS} vertex, and should be called as 
\begin{align*}
 {\tt CALL\ FVSIRX(RI,VC,SC,GR,FMASS,FWIDTH\ ,\ FVSIR)}.
\end{align*}
The output {\tt FVSIR} is a complex six-dimensional array:
\begin{multline}
 {\tt (FVSIR)}=
  \frac{i(\not\!p+m_{F})}{p^2-m_{F}^2+im_{F}\Gamma_{F}}\,{\tt SC(1)}\\
 \times [i{\tt GR(1)}^*P_{R}+i{\tt GR(2)}^*P_{L}]
  \gamma^{\mu}\!\not\!V({\tt RI})_{\mu}
\end{multline}
for the first four components of {\tt FVSIR(6)}, and
\begin{align}
 {\tt FVSIR(5)}&={\tt RI(17)-VC(5)-SC(2)},\\
 {\tt FVSIR(6)}&={\tt RI(18)-VC(6)-SC(3)},
\end{align}
for the momentum $p$.

\subsubsection{\tt JSIORX}

This subroutine computes an off-shell vector current {\tt J}
made from the interaction of a {\tt S}calar boson, a flowing-{\tt I}n
fermion and a flowing-{\tt O}ut {\tt R}-S fermion 
by the {\tt FRVS} vertex, and should be called as
\begin{center}
 {\tt CALL JSIORX(FI,RO,SC,GR,VMASS,VWIDTH , JSIOR)}.
\end{center}
What we compute here is 
\begin{multline}
 {\tt JSIOR(\nu+1)}=\frac{i}{q^2-m_{V}^2+im_{V}\Gamma_{V}}
  \left(-g^{\rho\nu}+\frac{q^{\rho}q^{\nu}}{m_{V}^{2}}\right)\\
 \times({\tt RO})_{\mu}{\tt SC(1)}\,\gamma_{\rho}\gamma^{\mu}
  [i{\tt GR(1)}P_{L}+i{\tt GR(2)}P_{R}]({\tt FI})
\end{multline}
for the massive vector boson, or
\begin{multline}
 {\tt JSIOR(\nu+1)}=\frac{-i}{q^2} \\
  \times({\tt RO})_{\mu}{\tt SC(1)}\,\gamma^{\nu}\gamma^{\mu}
   [i{\tt GR(1)}P_{L}+i{\tt GR(2)}P_{R}]({\tt FI})
\end{multline}
for the massless vector boson, and
\begin{align}
 {\tt JSIOR(5)}&={\tt -FI(5)+RO(17)+SC(2)},\\
 {\tt JSIOR(6)}&={\tt -FI(6)+RO(18)+SC(3)},
\end{align}
for the momentum $q$.

\subsubsection{\tt JSIROX}

This subroutine computes an off-shell vector current {\tt J} made from
the interaction of a {\tt S}calar boson, a flowing-{\tt I}n {\tt R}-S
fermion and a flowing-{\tt O}ut fermion
by the {\tt FRVS} vertex, and should be called as
\begin{center}
 {\tt CALL JSIROX(RI,FO,SC,GR,VMASS,VWIDTH , JSIRO)}.
\end{center}
What we compute here is
\begin{multline}
 {\tt JSIRO(\nu+1)}=\frac{i}{q^2-m_{V}^2+im_{V}\Gamma_{V}}
  \left(-g^{\rho\nu}+\frac{q^{\rho}q^{\nu}}{m_{V}^{2}}\right) \\
 \times({\tt FO})\,{\tt SC(1)}[i{\tt GR(1)}^*P_{R}+i{\tt GR(2)}^*P_{L}]
  \gamma^{\mu}\gamma_{\rho}({\tt RI})_{\mu}
\end{multline}
for the massive vector boson, or
\begin{multline}
 {\tt JSIRO(\nu+1)}=\frac{-i}{q^2} \\
 \times({\tt FO})\,{\tt SC(1)}[i{\tt GR(1)}^*P_{R}+i{\tt GR(2)}^*P_{L}]
  \gamma^{\mu}\gamma^{\nu}({\tt RI})_{\mu}
\end{multline}
for the massless vector boson, and 
\begin{align}
 {\tt JSIRO(5)}&={\tt -RI(17)+FO(5)+SC(2)},\\
 {\tt JSIRO(6)}&={\tt -RI(18)+FO(6)+SC(3)},
\end{align}
for the momentum $q$.

\subsubsection{\tt HVIORX}

This subroutine computes an off-shell scalar current {\tt H} made from
the interaction of a {\tt V}ector boson, a flowing-{\tt I}n fermion and a
flowing-{\tt O}ut {\tt R}-S fermion 
by the {\tt FRVS} vertex, and should be called as
\begin{align*}
 {\tt CALL\ HVIORX(FI,RO,VC,GR,SMASS,SWIDTH\ ,\ HVIOR)}.
\end{align*}
What we compute here is
\begin{multline}
 {\tt HVIOR(1)}=\frac{i}{q^2-m_{S}^2+im_{S}\Gamma_S} \\
 \times({\tt RO})_{\mu}\!\not\!V\gamma^{\mu}
  [i{\tt GR(1)}P_{L}+i{\tt GR(2)}P_{R}]({\tt FI}),
\end{multline}
and
\begin{align}
 {\tt HVIOR(2)}&={\tt -FI(5)+RO(17)+VC(5)},\\
 {\tt HVIOR(3)}&={\tt -FI(6)+RO(18)+VC(6)},
\end{align}
for the momentum $q$.

\subsubsection{\tt HVIROX}

This subroutine computes an off-shell scalar current {\tt H} made from 
the interaction of a {\tt V}ector boson, a flowing-{\tt I}n 
{\tt R}-S fermion and a flowing-{\tt O}ut fermion
by the {\tt FRVS} vertex, and should be called as
\begin{align*}
 {\tt CALL\ HVIROX(RI,FO,VC,GR,SMASS,SWIDTH\ ,\ HVIRO)}.
\end{align*}
What we compute here is
\begin{multline}
 {\tt HVIRO(1)}=\frac{i}{q^2-m_{S}^2+im_{S}\Gamma_S} \\
 \times({\tt FO})[i{\tt GR(1)}^*P_{R}+i{\tt GR(2)}^*P_{L}]
  \gamma^{\mu}\!\not\!V({\tt RI})_{\mu},
\end{multline}
and
\begin{align}
 {\tt HVIRO(2)}&={\tt -RI(17)+FO(5)+VC(5)},\\
 {\tt HVIRO(3)}&={\tt -RI(18)+FO(6)+VC(6)},
\end{align}
for the momentum $q$.

%%%%%%%%%%%%%%%%%%%%%%%%%%%%%%%%%%%%%%%%%%%%%%%%%%%%%%%%%%%%%%%%%%%%
\subsection{FRVV vertex}\label{sec:vertex_f}

The {\tt FRVV} vertices are obtained from the interaction Lagrangian
among a fermion, a R-S fermion and two vector bosons:
\begin{multline}
 {\cal L}_{\tt FRVV}=if^{abc}
 \overline{R}_{\mu}^{}[\gamma^{\nu},\gamma^{\rho}]\gamma^{\mu}
 [{\tt GR(1)}P_L+{\tt GR(2)}P_R]f^{a}\,V_{\nu}^{b}V_{\rho}^{c}\\
 +{\rm h.c.}
\end{multline}
with the structure constant $f^{abc}$, which can be handled by the 
{\tt MG} automatically. 
The coupling constant {\tt GR} is the product of 
the {\tt FRV} coupling constant
and the gauge coupling constant of the involving gauge boson
as in the {\tt FRVS} coupling; see \eqref{GR_GFRVS}.

\subsubsection{\tt IORVVX}

This subroutine computes an amplitude of the {\tt FRVV} vertex from a
flowing-{\tt I}n fermion, a flowing-{\tt O}ut {\tt R}-S fermion and two
{\tt V}ector bosons, and should be called as
\begin{center}
 {\tt CALL IORVVX(FI,RO,VA,VB,GR , VERTEX)}.
\end{center}
What we compute here is
\begin{align}
 {\tt VERTEX}=({\tt RO})_{\mu}[\not\!V^a,\not\!V^b]\gamma^{\mu}
  [{\tt GR(1)}P_{L}+{\tt GR(2)}P_{R}]({\tt FI}), 
\end{align}
where we use the notations
\begin{align}
 V^{a,\mu}&={\tt VA}(\mu+1),\\
 V^{b,\mu}&={\tt VB}(\mu+1).
\end{align}

\subsubsection{\tt IROVVX}

This subroutine computes an amplitude of the {\tt FRVV} vertex from a
flowing-{\tt I}n {\tt R}-S fermion, a flowing-{\tt O}ut fermion and two
{\tt V}ector bosons, and should be called as
\begin{center}
 {\tt CALL IROVVX(RI,FO,VA,VB,GR , VERTEX)}.
\end{center}
What we compute here is
\begin{align}
 {\tt VERTEX}=({\tt FO})
  [{\tt GR(1)}^*P_{R}+{\tt GR(2)}^*P_{L}]\gamma^{\mu}
  [\not\!V^a,\not\!V^b]({\tt RI})_{\mu}.
\end{align}

\subsubsection{\tt FVVORX}

This subroutine computes an off-shell {\tt F}ermion wavefunction made
from the interaction of two {\tt V}ector bosons and a flowing-{\tt O}ut
{\tt R}-S fermion by the {\tt FRVV} vertex, and should be called as
\begin{align*}
 {\tt CALL\ FVVORX(RO,VA,VB,GR,FMASS,FWIDTH\ ,\ FVVOR)}.
\end{align*}
What we compute here is
\begin{multline}
 {\tt (FVVOR)}=({\tt RO})_{\mu}
  [\not\!V^a,\not\!V^b]\gamma^{\mu}[i{\tt GR(1)}P_{L}+i{\tt GR(2)}P_{R}] \\
\times\frac{i(\not\!p+m_{F})}{p^2-m_{F}^2+im_{F}\Gamma_{F}},
\end{multline}
and
\begin{align}
 {\tt FVVOR(5)}&={\tt RO(17)+VA(5)+VB(5)},\\
 {\tt FVVOR(6)}&={\tt RO(18)+VA(6)+VB(6)}.
\end{align}

\subsubsection{\tt FVVIRX}

This subroutine computes an off-shell {\tt F}ermion wavefunction made
from the interaction of two {\tt V}ector bosons and a flowing-{\tt I}n
{\tt R}-S fermion by the {\tt FRVV} vertex, and should be called as
\begin{align*}
 {\tt CALL\ FVVIRX(RI,VA,VB,GR,FMASS,FWIDTH\ ,\ FVVIR)}.
\end{align*}
What we compute here is
\begin{multline}
 {\tt (FVVIR)}=
  \frac{i(\not\!p+m_{F})}{p^2-m_{F}^2+im_{F}\Gamma_{F}} \\
 \times[i{\tt GR(1)}^*P_{R}+i{\tt GR(2)}^*P_{L}]\gamma^{\mu}
  [\not\!V^a,\not\!V^b]({\tt RI})_{\mu},
\end{multline}
and
\begin{align}
 {\tt FVVIR(5)}&={\tt RI(17)-VA(5)-VB(5)},\\
 {\tt FVVIR(6)}&={\tt RI(18)-VA(6)-VB(6)}.
\end{align}

\subsubsection{\tt JVIORX}

This subroutine computes an off-shell vector current {\tt J} made from 
the interaction of a {\tt V}ector boson, a flowing-{\tt I}n fermion and
a flowing-{\tt O}ut {\tt R}-S fermion by the {\tt FRVV} vertex, and
should be called as
\begin{center}
 {\tt CALL JVIORX(FI,RO,VC,GR,VMASS,VWIDTH , JVIOR)}.
\end{center}
What we compute here is
\begin{multline}
 {\tt JVIOR(\nu+1)}=\frac{i}{q^2-m_{V}^2+im_{V}\Gamma_{V}}
  \left(-g^{\rho\nu}+\frac{q^{\rho}q^{\nu}}{m_{V}^{2}}\right) \\
 \times({\tt RO})_{\mu}[\gamma_{\rho},\not\!V]\gamma^{\mu}
 [i{\tt GR(1)}P_{L}+i{\tt GR(2)}P_{R}]({\tt FI})
\end{multline}
for the massive vector boson, or
\begin{multline}
 {\tt JVIOR(\nu+1)}=\frac{-i}{q^2} \\
 \times({\tt RO})_{\mu}[\gamma^{\nu},\not\!V]\gamma^{\mu}
  [i{\tt GR(1)}P_{L}+i{\tt GR(2)}P_{R}]({\tt FI})
\end{multline}
for the massless vector boson, and
\begin{align}
 {\tt JVIOR(5)}&={\tt -FI(5)+RO(17)+VC(5)},\\
 {\tt JVIOR(6)}&={\tt -FI(6)+RO(18)+VC(6)}.
\end{align}

\subsubsection{\tt JVIROX}

This subroutine computes an off-shell vector current {\tt J} made from
the interaction of a {\tt V}ector boson, a flowing-{\tt I}n {\tt R}-S
fermion and a flowing-{\tt O}ut fermion by the {\tt FRVV} vertex, and
should be called as
\begin{center}
 {\tt CALL JVIROX(RI,FO,VC,GR,VMASS,VWIDTH , JVIRO)}.
\end{center}
What we compute here is
\begin{multline}
 {\tt JVIRO(\nu+1)}=\frac{i}{q^2-m_{V}^2+im_{V}\Gamma_{V}}
  \left(-g^{\rho\nu}+\frac{q^{\rho}q^{\nu}}{m_{V}^{2}}\right) \\
 \times({\tt FO})[i{\tt GR(1)}^*P_{R}+i{\tt GR(2)}^*P_{L}]
  \gamma^{\mu}[\gamma_{\rho},\not\!V]({\tt RI})_{\mu}
\end{multline}
for the massive vector boson, or
\begin{multline}
 {\tt JVIRO(\nu+1)}=\frac{-i}{q^2} \\
 \times({\tt FO})[i{\tt GR(1)}^*P_{R}+i{\tt GR(2)}^*P_{L}]
  \gamma^{\mu}[\gamma^{\nu},\not\!V]({\tt RI})_{\mu}
\end{multline}
for the massless vector boson, and
\begin{align}
 {\tt JVIRO(5)}&={\tt -RI(17)+FO(5)+VC(5)},\\
 {\tt JVIRO(6)}&={\tt -RI(18)+FO(6)+VC(6)}.
\end{align}
%

%%%%%%%%%%%%%%%%%%%%%%%%%%%%%%%%%%%%%%%%%%%%%%%%%%%%%%%%%%%%%%%%%%%%
\subsection{Checking for the new HELAS subroutines}\label{sec:check}

The new {\tt HELAS} subroutines are tested by using the gauge
invariance of the
helicity amplitudes. In particular, we use the following processes; 
\begin{align*}
 &qg\to\tilde q\tilde G
  &{\rm for}\ &{\tt IORSXX}, {\tt IROSXX}, {\tt FSORXX}, {\tt FSIRXX},\\ 
  &&&          {\tt HIORXX}, {\tt HIROXX}, {\tt IORVSX}, {\tt IROVSX},\\
 &gg\to\tilde g\tilde G
  &{\rm for}\ &{\tt IORVXX}, {\tt IROVXX}, {\tt FVORXX}, {\tt FVIRXX},\\ 
  &&&          {\tt JIORXX}, {\tt JIROXX}, {\tt IORVVX}, {\tt IROVVX},\\
 &qg\to\tilde q\tilde Gg
  &{\rm for}\ &{\tt FVSORX}, {\tt FVSIRX}, {\tt JSIORX}, {\tt JSIROX},\\
  &&&          {\tt HVIORX}, {\tt HVIROX},\\
 &gg\to\tilde g\tilde Gg
  &{\rm for}\ &{\tt FVVORX}, {\tt FVVIRX}, {\tt JVIORX}, {\tt JVIROX}.
\end{align*}
More explicitly, we express the helicity amplitudes of the above
processes as 
\begin{align} 
 {\cal M}_{\lambda_{\tilde G}\lambda_g} = 
 \bar\psi_{\mu}(p_{\tilde G},\lambda_{\tilde G})\,
 T^{\mu\nu}
 \epsilon_\nu(p_g,\lambda_g)
\end{align}
or
\begin{align} 
 {\cal M}_{\lambda_{\tilde G}\lambda_g} =  
 T^{\mu\nu}\,
 \psi_{\mu}(p_{\tilde G},\lambda_{\tilde G})\,
 \epsilon_\nu(p_g,\lambda_g)
\end{align}
with an external spin-3/2 and a gluon wavefunction. 
The identity for the $SU(3)$ gauge invariance
\begin{align}
 {p_g}_{\nu}\,T^{\mu\nu}=0
\end{align}
tests all the above subroutines thoroughly.
We also test the agreement of the helicity-summed squared amplitudes
at arbitrary Lorentz frames.

%%%%%%%%%%%%%% Begin App. 2 %%%%%%%%%%%%%%%%%%%%%%%%%%%%%%%%%%%%%% 
\section{Implementation of spin-3/2 gravitinos into MadGraph}
\label{sec:mg}

\begin{table}[b]
\centering
\begin{tabular}{ccllllccl}
\hline\hline
\multicolumn{7}{l}{3-point couplings}  & {\tt GR}  & \\
\hline
 FRS && {\tt q} & {\tt gro} & {\tt ql} &&& {\tt GFRSL} & \\
     && {\tt q} & {\tt gro} & {\tt qr} &&& {\tt GFRSR} & \\
 FRV && {\tt go} & {\tt gro} & {\tt g} &&& {\tt GFRV} & \\
\hline\hline
\multicolumn{7}{l}{4-point couplings}        &{\tt GR}   &  \\
\hline
 FRVS &&{\tt q} &{\tt gro} &{\tt g} &{\tt ql} &&{\tt GFRGSL} &\!\!\!\!\!= {\tt GFRSL*GG} \\
      &&{\tt q} &{\tt gro} &{\tt g} &{\tt qr} &&{\tt GFRGSR} &\!\!\!\!\!= {\tt GFRSR*GG} \\
 FRVV &&{\tt go} &{\tt gro} &{\tt g} &{\tt g} &&{\tt GGORGG} &\!\!\!\!\!= {\tt GFRV*G} \\
\hline
\end{tabular}
\caption{List of the coupling constants for each gravitino vertex involving  
SUSY QCD particles. All the
 particles and the coupling constants are written in the {\tt MG}
 notation. {\tt gro} stands for a massive gravitino, 
 {\tt q} represents a light quark, 
 and {\tt ql}/{\tt qr} is a left/right-handed squark.
 {\tt g} and {\tt go} are a gluon and a gluino, respectively.
 {\tt GR} is a non-renormalizable coupling
 constant defined in each subroutine in App.~\ref{sec:helas_new}.}
\label{couplist}
\end{table}

In this appendix, we describe how we implement spin-3/2
gravitinos and their interactions into {\tt MG}.

First, using the default {\tt mssm} model in 
{\tt MG/ME\,v4}~\cite{Alwall:2007st}, we make
our new model directory, {\tt mssm\_gravitino}, 
including a massive gravitino ({\tt particles.dat}) and
its interactions with 
SM and SUSY particles ({\tt interactions.dat} and {\tt couplings.f});
we show the coupling constants for each gravitino vertex involving 
SUSY QCD particles in Table~\ref{couplist} as examples.
Then we add all the new {\tt HELAS} subroutines
for spin-3/2 gravitinos to the {\tt HELAS} library in {\tt MG}.
Since the present {\tt MG} does not handle spin-3/2 particles, we
further modify the codes in {\tt MG} to tell it how to generate the 
{\tt FRS}, {\tt FRV}, {\tt FRVS} and {\tt FRVV} type of vertices and helicity
amplitudes, and how to
deal with the helicity of external spin-3/2 particles.

%RRRRRRRRRRRRRRRRRRRRRRRRRRRRRRRRRRRRRRRRRRRRRRRRRRRRRRRRRRRRR

\end{document}